\documentclass[showpacs,preprintnumbers,amsmath,amssymb,
superscriptaddress,floatfix]{revtex4}
\usepackage{graphicx}
\usepackage{amsmath}
\usepackage{bm}
\usepackage{mathrsfs}
\usepackage{color}
\usepackage{slashed}
\usepackage{dcolumn}
\usepackage[normalem]{ulem}

\newcommand{\be}{\begin{equation}}
\newcommand{\ee}{\end{equation}}
\newcommand{\ba}{\begin{eqnarray}}
\newcommand{\ea}{\end{eqnarray}}

\newcommand{\pr}{\prime}

\begin{document}

\title{The possible $K \bar{K}^*$ and $D \bar{D}^*$ bound and resonance states by solving Schrodinger equation}

\author{Bao-Xi Sun}
\email{sunbx@bjut.edu.cn}
\affiliation{School of Physics and Optoelectronic Engineering, Beijing University of Technology, Beijing 100124, China}

\author{Qin-Qin Cao}
\email{s202166092@emails.bjut.edu.cn}
\affiliation{School of Physics and Optoelectronic Engineering, Beijing University of Technology, Beijing 100124, China}

\author{Ying-Tai Sun}
\email{3071450876@qq.com}
\affiliation{School of Mechanical and Materials Engineering, North China University of Technology, Beijing 100144, China}

\date{\today}

\begin{abstract}
The Schrodinger equation with a Yukawa type of potential is solved analytically. When different boundary conditions are taken into account, a series of solutions are indicated as  Bessel function, the first kind of Hankel function and the second kind of Hankel function, respectively. Subsequently, the scattering processes of $K \bar{K}^*$ and $D \bar{ D}^*$ are investigated.
In the $K \bar{K}^*$ sector, the $f_1(1285)$ particle is treated as a $K \bar{K}^*$ bound state, therefore, the coupling constant in the $K \bar{K}^*$ Yukawa potential can be fixed according to the binding energy of the $f_1(1285)$ particle. Consequently,  a $K \bar{K}^*$ resonance state is generated by solving the Schrodinger equation with the outgoing wave condition, which lie at $1417-i18$MeV on the complex energy plane. It is reasonable to assume  that the $K \bar{K}^*$ resonance state at $1417-i18$MeV might correspond to the $f_1(1420)$ particle in the review of Particle Data Group(PDG).
In the $D \bar{D}^*$ sector, since the $X(3872)$ particle is almost located at the $D \bar{ D}^*$ threshold, the binding energy of it equals to zero approximately. Therefore, the coupling constant in the $D \bar{ D}^*$ Yukawa potential is determined, which is related to the first zero point of the zero order Bessel function. Similarly to the $K \bar{K}^*$ case, four resonance states are produced as solutions of the Schrodinger equation with the outgoing wave condition. It is assumed that the resonance states at $3885-i1$MeV, $4029-i108$ MeV, $4328-i191$MeV and $4772-i267$MeV might be associated with the $Zc(3900)$, the $X(3940)$, the $\chi_{c1}(4274)$ and $\chi_{c1}(4685)$ particles, respectively. It is noted that all solutions are isospin degenerate.
\end{abstract}

%
%

\maketitle

\section{Introduction}
\label{sect:Introduction}

The Schrodinger equation in a central force potential can be solved with the method of separation of variables, and the eigenenergy is only relevant to the radial part of the Schrodinger equation. In the S-wave approximation, where the quantum number of the orbital angular momentum is zero,  the centrifugal potential term disappears, and the radial Schrodinger equation takes the same  form as the one-dimensional Schrodinger equation with appropriate  function substitution.

If the wave function vanishes at the infinity, the solution of the Schrodinger equation corresponds to a bound state of the system. This is a typical problem in the textbook of quantum mechanics. However, here we will try to study a scattering process  that the particle comes into the potential from the infinity, or goes out of the potential directly, which implies that the wave function will not disappear at the infinity. In this situation, complex eigenenergies of the Hamiltonian are obtained when the Schrodinger equation is solved. Actually, these solutions are associated with the complex poles of the scattering matrix, and they correspond to different types of resonance states, respectively\cite{Moiseyev}.

The $f_1(1285)$ and $f_1(1420)$ particles are assumed to be quark-antiquark states in a three-flavor linear sigma model although their masses are above 1GeV\cite{Rischke}.
Meanwhile, the $f_1(1285)$ particle has been studied in the unitary coupled-channel approximation by solving the Bethe-Salpeter equation, and it is asserted that the $f_1(1285)$ particle should be a $K\bar{K}^*$ or $K^* \bar{K}$ bound state since its mass is lower than the $K\bar{K}^*$ threshold\cite{Roca2005}, while the $f_1(1420)$ particle is related to a triangle singularity of $K^* \bar{K}K$\cite{oset1420}.
However, in Ref.~\cite{sun1420}, where the longitudinal part of the vector meson propagator is taken into account in the intermediate loop function when the Bethe-Salpeter equation is solved, a peak appears in the vicinity of 1400MeV, which is above the $K\bar{K}^*$ threshold, and no other peaks are detected. Therefore, it is assumed that this peak might correspond a $K\bar{K}^*$ resonance state and is identified with the $f_1(1420)$ particle. Apparently,  these two articles show different results from each other.

The proton-neutron resonance states have been obtained by solving the Schrodinger equation under the outgoing wave condition\cite{Sunbxinpreparation}, where an one-pion-exchanging potential is assumed, just as done in Re.~\cite{Zhangyongde}.
In this work, the $K \bar{K}^*$ interaction is investigated by solving the Schrodinger equation under the outgoing wave condition.  An one-pion-exchange potential in the $K \bar{K}^*$ system is assumed, which is different from the kernel used in the unitary coupled-channel approximation, where a vector meson exchange potential is dominant according to the SU(3) hidden gauge symmetry.
We assume the $ f_1(1285)$ particle is a $K \bar{K}^*$ bound state, and then the $K \bar{K}^*$ coupling constant is determined. Sequently, a $K \bar{K}^*$ resonance state around 1400MeV is obtained as a solution of the Schrodinger equation under the outgoing wave condition. Apparently, it's more possible that the $K \bar{K}^*$ resonance state in the vicinity of 1400MeV might correspond to the $ f_1(1420)$ particle in the review of Particle Data Group. Therefore, the relation between the $K \bar{K}^*$ bound state $ f_1(1285)$ and the $K \bar{K}^*$ resonance state $ f_1(1420)$ is established. At this point, it is different from those calculation results by solving the Bethe-Salpeter equation, where only one pole of the $ T-$ amplitude is generated dynamically on the complex energy plane.

In sequence, this method is extended to study the $D \bar{ D}^*$  system reasonably by replacing the $ s-$ quark into the  $ c-$ quark.
Ever since the $X(3872)$ particle was discovered by Belle collaboration firstly in 2003\cite{X3872-2003}, more charmonia have been found in the facilities around the world, and more detailed iterations can be found in recent review articles\cite{Yangzhi1,Yangzhi2,Yangzhi3,Yangzhi4,Yangzhi5,Yangzhi6,Yangzhi7,ZhiGangWang}.
On the structure of the $X(3872)$ with $J^{PC}=1^{++}$ (also named as $\chi_{c1}(3872)$ in Ref.~\cite{PDG}), there are different theoretical interpretations, such as the conventional twisted $\chi_{c1}(2P)$ charmonium\cite{june10,june11}, the compact tetraquark state\cite{june12,june13,sun29,sun30}, the hybrid state\cite{sun31}, the $D \bar{D}^*/ D^* \bar{D}$ bound state\cite{june14,june15,june16,june17,sun22,sun23,sun24,sun25,SunX3872}, the virtual state of $D \bar{D}^*/ D^* \bar{D}$\cite{sun26,sun27}, the mixture of $c\bar{c}$ and $D \bar{D}^*/ D^* \bar{D}$ bound state\cite{june18,june19,june20,june21,june22,june23,june24,KangXianWei}. Especially, the $X(3872)$ particle is also studied with the method of pole counting rule, and it is concluded that two nearby poles are essential to describe the experimental data\cite{sun36,sun37}.

In the present work, the $K \bar{K}^*$ and $D \bar{D}^*$ systems are studied respectively by solving Schrodinger equation analytically in the one-pion exchanging potential, and some resonance states are obtained, and more of them have a counterpart in the review of Particle Data Group(PDG)\cite{PDG}.

The whole article is organized as follows. In Section~\ref{sect:framework}, the framework is evaluated in detail. the $K \bar{K}^*$ and $D \bar{D}^*$ systems are analyzed in Sections~\ref{sect:KKstar}~and~\ref{sect:DDstar}, respectively. Finally, a summary is given in Section~\ref{sect:summary}.

\section{The Schrodinger equation with a Yukawa potential}
\label{sect:framework}

If the interaction of two particles is realized by exchanging a pion, the potential of them can be indicated as a Yukawa type, i.e.,
\be
\label{eq:202307071816}
V(r)=-g^2\frac{e^{-mr}}{d},
\ee
where $m$ is the mass of the pion, $g$ is the coupling constant, and the distance $r$ in the denominator has been  replaced with the range of force $d=1/m$ approximately.
It is apparent that the potential in Eq.~(\ref{eq:202307071816}) is reasonable in the range of force, and it is equal to the original Yukawa potential asymptotically at the infinity. Under this approximation, the Schrodinger equation can be solved analytically.

Supposing the radial wave function $R(r)=\frac{u(r)}{r}$, the radial Schrodinger equation with $l=0$ can be written as
\be
\label{eq:202307081218}
-\frac{\hbar^2}{2\mu} \frac{d^2 u(r)}{dr^2}+V(r)u(r)=Eu(r),
\ee
where $\mu$ is the reduced mass of the two-body system.

With the variable substitution
\be
r \rightarrow x=\alpha e^{-\beta r},~~~~0 \le x \le \alpha,
\ee
and
\be
u(r)=J(x),
\ee
the radial Schrodinger equation in Eq.~(\ref{eq:202307081218}) becomes
\be
\frac{d^2 J(x)}{d x^2}+\frac{1}{x} \frac{d J(x)}{d x} +\left[\frac{2\mu g^2}{d \beta^2} \frac{x}{\alpha}^{\frac{1}{d \beta}} \frac{1}{x^2}+\frac{2 \mu E}{\beta^2} \frac{1}{x^2} \right] J(x)=0.
\ee
Supposing
\be
\alpha=2g \sqrt{2 \mu d},~~~~\beta=\frac{1}{2d},
\ee
and
\be
\label{eq:rhoenergy}
\rho^2=-8d^2 \mu E,~~~~E \le 0,
\ee
the radial Schrodinger equation becomes the $\rho$th order Bessel equation, i.e.,
\be
\label{eq:202307081903}
\frac{d^2 J(x)}{d x^2}+\frac{1}{x} \frac{d J(x)}{d x} +\left[1-\frac{\rho^2}{x^2}\right] J(x)=0,
\ee
and its solution is the $\rho$th order Bessel function $J_\rho(x)$.

For the bound state, when $r \rightarrow +\infty$, the radial wave function $R(r) \rightarrow 0$, which implies that $u(r)=J_\rho(\alpha e^{-\beta r})=J_\rho(0)$ with $\rho \ge 0$. On the other hand, when $r \rightarrow 0$, $u(r) \rightarrow 0$, and it means
\be
\label{eq:Besselzeropoint}
J_\rho(\alpha)=0.
\ee
Therefore, if only one bound state of the two-body system has been detected and the binding energy is given, the order of the Bessel function $\rho$ in Eq.~(\ref{eq:Besselzeropoint}) can be determined according to Eq.~(\ref{eq:rhoenergy}), and then the coupling constant $g$ in the Yukawa potential is obtained with the first zero point of the Bessel function $J_\rho(\alpha)$, which takes a form of
\be
\label{eq:couplingzeropoint}
g^2=\frac{\alpha^2}{8 \mu  d}.
\ee

The Hankel functions $H_\rho^{(1)}(x)$ and $H_\rho^{(2)}(x)$ are also two independent solutions of the Bessel equation. Actually, $H_\rho^{(1)}(x) e^{-i \omega t}$ represents a wave along the positive direction of the $x$ axis, while $H_\rho^{(2)}(x) e^{-i \omega t}$ corresponds to a wave along the negative direction of the $x$ axis. When a scattering process of two particles is investigated, the general solution of Eq.~(\ref{eq:202307081903}) can be written as
\be
\label{eq:202307081915}
u(r)=D H_\rho^{(1)}(x) +D^\pr H_\rho^{(2)}(x).
\ee
By requiring $D^\pr=0$, only the outgoing wave $H_\rho^{(1)}(x)$ is left. Similarly, with $D=0$, the incoming wave $H_\rho^{(2)}(x)$ is conserved.

When the coefficient $D^\pr$ is set to zero in Eq.~(\ref{eq:202307081915}), $D^\pr=0$, the first kind of Hankel function $H_\rho^{(1)}(x)$ represents a wave coming in from $r=\infty$. When $r\rightarrow +\infty$,
$u(r) \sim H_\rho^{(1)}(\alpha e^{-\beta r})\rightarrow H_\rho^{(1)}(0)$. when $r\rightarrow0$, $u(0)\rightarrow0$, which implies
\be
\label{eq:incomingwavecon}
H_\rho^{(1)}(\alpha)=0.
\ee
Actually, the incoming wave condition in Eq.~(\ref{eq:incomingwavecon}) is equivalent to the bound wave condition in Eq.~(\ref{eq:Besselzeropoint}) since the zero points of $H_\rho^{(1)}(\alpha)$ are the same as those of $J_\rho(\alpha)$ respectively when the value of $\rho$ is determined. Therefore, either by Eq.~(\ref{eq:Besselzeropoint}) or by Eq.~(\ref{eq:incomingwavecon}), the same value of the coupling constant $g$ in the Yukawa potential can be obtained with the fixed binding energy.

When the coefficient $D$ is set to zero in Eq.~(\ref{eq:202307081915}), $D=0$, the second kind of Hankel function $H_\rho^{(2)}(x)$ represents a wave going out from the coordinate origin $r=0$. When $r\rightarrow+\infty$,
$u(r) \sim H_\rho^{(2)}(\alpha e^{-\beta r})\rightarrow H_\rho^{(2)}(0)$. when $r \rightarrow 0$, $u(0) \rightarrow 0$, which implies
\be
\label{eq:outgoing}
H_\rho^{(2)}(\alpha)=0.
\ee
If the first zero point of the Bessel function $J_\rho(\alpha)$ has been obtained according to the binding energy of the two-body system, which indicates the coupling constant $g$ in the Yukawa potential is determined, the energies of the two-body resonance states  can be calculated with the outgoing wave condition in Eq.~(\ref{eq:outgoing}). Apparently, the energy of the resonance state is related to the order of the second kind of Hankel function, and takes a complex value
$E=M-i\frac{\Gamma}{2}$, where the real part represents the mass of the resonance state, while the imaginary part is one half of the decay width, i.e., $\Gamma=-2iImE$, as discussed in Ref.~\cite{Moiseyev}.

\section{$K \bar{K}^*$ interaction}
\label{sect:KKstar}

\begin{figure}
\includegraphics[width=0.8\textwidth]{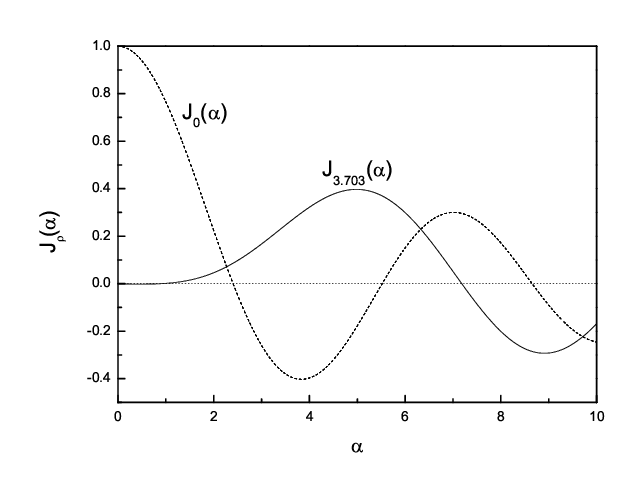}
\caption{
The Bessel function $J_\rho(\alpha)$ with $\rho=3.703$ for the $K \bar{K}^*$ system and the first nonzero zero-point lies at $\alpha=7.1831$, which is assumed to correspond to the $f_1(1285)$ particle. The Bessel function $J_0(\alpha)$ is also depicted in the figure.
}
\label{fig:Bessel}
\end{figure}

Hence we try to study the possible bound or resonance states of the $K\bar{K}^*$ system by solving the Schrodinger equation,  just as done in the case of the proton-neutron system\cite{Sunbxinpreparation}. The potential of the kaon( anyikaon) and the vector antikaon(kaon) takes the Yukawa type as that in Eq.~(\ref{eq:202307071816}), where the range of force is the reciprocal of the pion mass, i.e., $d=\frac{1}{m}$ with $m=139.57$MeV. Since the mass of the $f_1(1285)$ particle is 105MeV lower than the $K\bar{K}^*$ threshold, it can be regarded as a $K\bar{K}^*$ bound state. Similarly to the case of the proton-neutron system, the order of Bessel function in Eq.~(\ref{eq:Besselzeropoint}) can be obtained with the binding energy $E=105$MeV, which takes a value of $3.703$, and then the first zero point of $J_\rho(\alpha)$ is  found to be $\alpha=7.1831$, as shown in Fig.~\ref{fig:Bessel}. Supposing the $f_1(1285)$ particle is a bound state of the $K\bar{K}^*$ system, the coupling constant of the $K\bar{K}^*$ potential $g$ is assumed to be relevant to the first zero point of Bessel function. Thus the coupling constant $g$ in the $K\bar{K}^*$ potential can be determined according to Eq.~(\ref{eq:couplingzeropoint}), i.e., $g=1.682$.

In follows, with the same value of $\alpha$ being the zero point, the order of the second kind of Hankel function will be evaluated according to Eq.~(\ref{eq:outgoing}). The order of the second kind of Hankel function might be complex, which implies that the complex eigenvalue of the Hamiltonian might correspond to a resonance state of the system. As to the $K\bar{K}^*$ system, the corresponding eigenenergy is listed in Table~\ref{table:KKstar}. Note that the value of $K\bar{K}^*$ threshold has been included in the real part of the energy.

\begin{figure}
\includegraphics[width=0.8\textwidth]{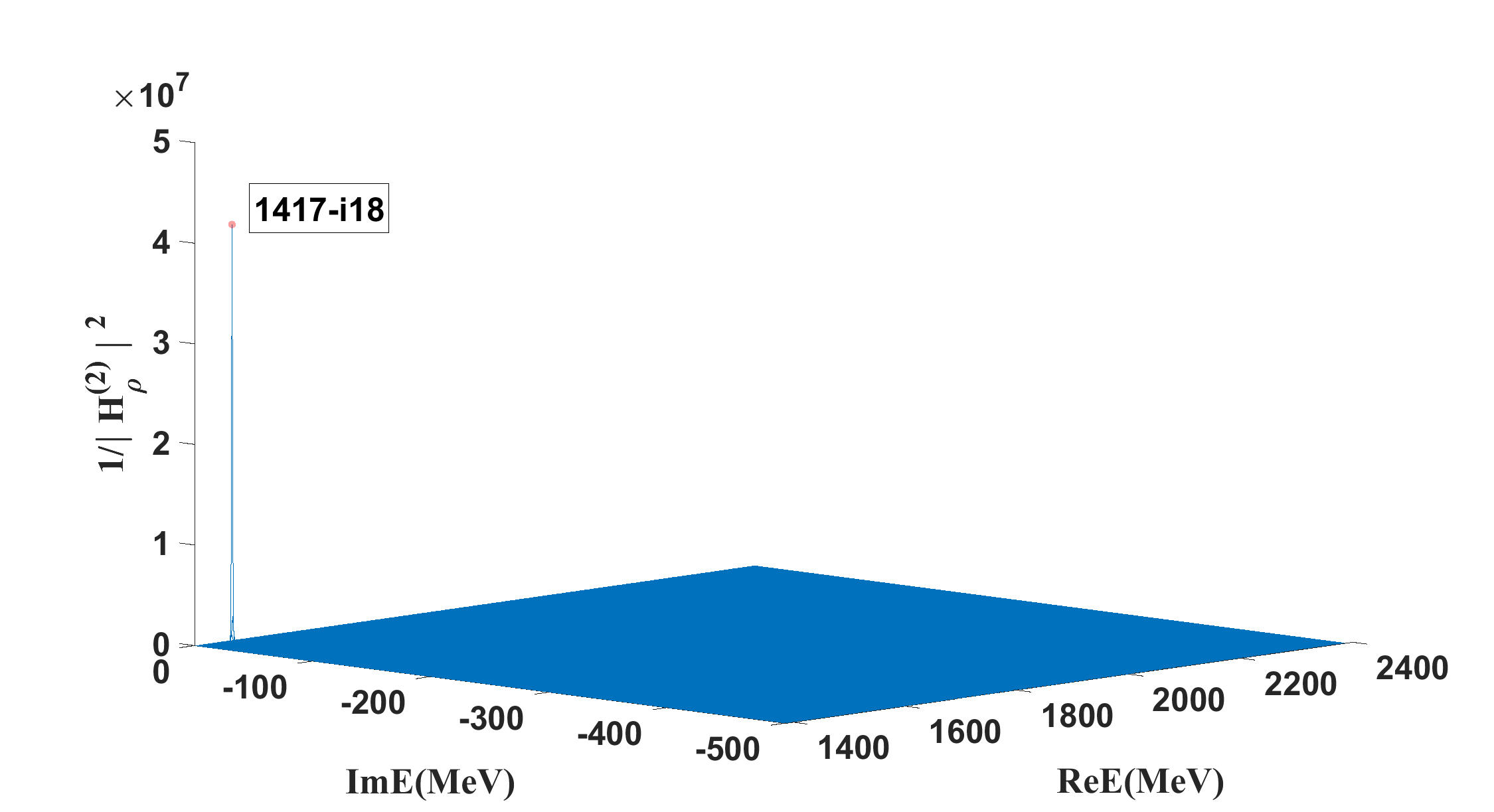}
\caption{$1/|H_\rho^{(2)}(\alpha)|^2$ .vs. the complex energy $E$ with $\alpha=7.1831$ in the $K \bar{K}^*$ case. The pole of $1/|H_\rho^{(2)}(\alpha)|^2$ corresponds to a zero-point of the second kind of Hankel function $H_\rho^{(2)}(\alpha)$, which represents a $K \bar{K}^*$ resonance state, as labeled in the figure.}
\label{fig:kkstar}
\end{figure}

The resonance state appears at $1417-i18$MeV, as labeled in Fig.~\ref{fig:kkstar}, where the function of $1/|H_\rho^{(2)}(\alpha)|^2$ is calculated at different complex energies. Apparently, a pole of $1/|H_\rho^{(2)}(\alpha)|^2$ corresponds to a zero point of $H_\rho^{(2)}(\alpha)$.
It is higher than the $K\bar{K}^*$ threshold, and might correspond to the $f_1(1420)$ particle.
Therefore, by solving the Schrodinger equation with different  boundary condition of the wave function at $r \rightarrow 0 $,   the $f_1(1285)$ and $f_1(1420)$ particles are generated in the S-wave approximation, which can be regarded as a bound state and a resonance state of the $K\bar{K}^*$ system, respectively.
In a word, it is assumed that the $K\bar{K}^*$ interaction is realized by exchanging a pion, which is different from the assertion of hidden gauge symmetry, where the flavor $SU(3)$ symmetry of hadrons is breaking spontaneously and a vector meson is assumed to transfer between the kaon and the vector anti-kaon, such as $\rho$, $\omega$ or $\phi$\cite{sun1420}.
\begin{table}[htbp]
 \renewcommand{\arraystretch}{1.2}
\centering
\vspace{0.5cm}
\begin{tabular}{c|c|c|c|c|c}
\hline\hline
 $K \bar{K}^*$ & energy &   name & $I^G(J^{PC})$ &  mass & width \\
 \hline
1 & $1417-i18$ & $f_1(1420)$ & $0^+(1^{++})$ &   $1426.3\pm0.9$  & $54.5\pm2.6$
  \\
\hline\hline
\end{tabular}
\caption{
The complex energy of the possible $K \bar{K}^*$ ($\bar{K} K^*$) resonance state with the outgoing wave condition in Eq.~(\ref{eq:outgoing}) and the possible counterpart in the PDG data. All are in units of MeV.
}\label{table:KKstar}
\end{table}

\section{$D \bar{D}^*$ Interaction}
\label{sect:DDstar}

In this section, we study the properties of charmonia by solving the Schrodinger equation of the $D$ and $\bar{D}^*$ mesons.
Supposing that the $D \bar{D}^*$ potential takes a Yukawa type, as given in Eq.~(\ref{eq:202307071816}), where $m=139.57$MeV, the $D \bar{D}^*$ interaction is realized by exchanging a pion. In order to determine the coupling constant $g$, the $X(3872)$ particle is assumed to be a $D \bar{D}^*$ bound state. Since the mass of $X(3872)$ particle almost lies at the $D \bar{D}^*$ threshold, the order of Bessel function is zero according to Eq.~(\ref{eq:rhoenergy}). Therefore, the value of the $D \bar{D}^*$ coupling constant is related to the first zero point of Bessel function $J_0(\alpha)$, which is $2.405$, thus the coupling constant in the $D \bar{D}^*$ potential can be obtained according to Eq.~(\ref{eq:couplingzeropoint}), i.e.,
\be
\label{eq:20230924}
g=0.323.
\ee
With  $\alpha=2.405$ as a zero point, the order of the second kind of Hankel function  can be obtained by solving Eq.~(\ref{eq:outgoing}), which is a complex number, and is relevant to the energy  and decay width of the $D \bar{D}^*$ resonance state. The corresponding energies of the $D \bar{D}^*$ resonance states are listed in Table~\ref{Table:DDstar}, moreover, the possible PDG counterparts are also depicted correspondingly.

\begin{figure}
\includegraphics[width=0.8\textwidth]{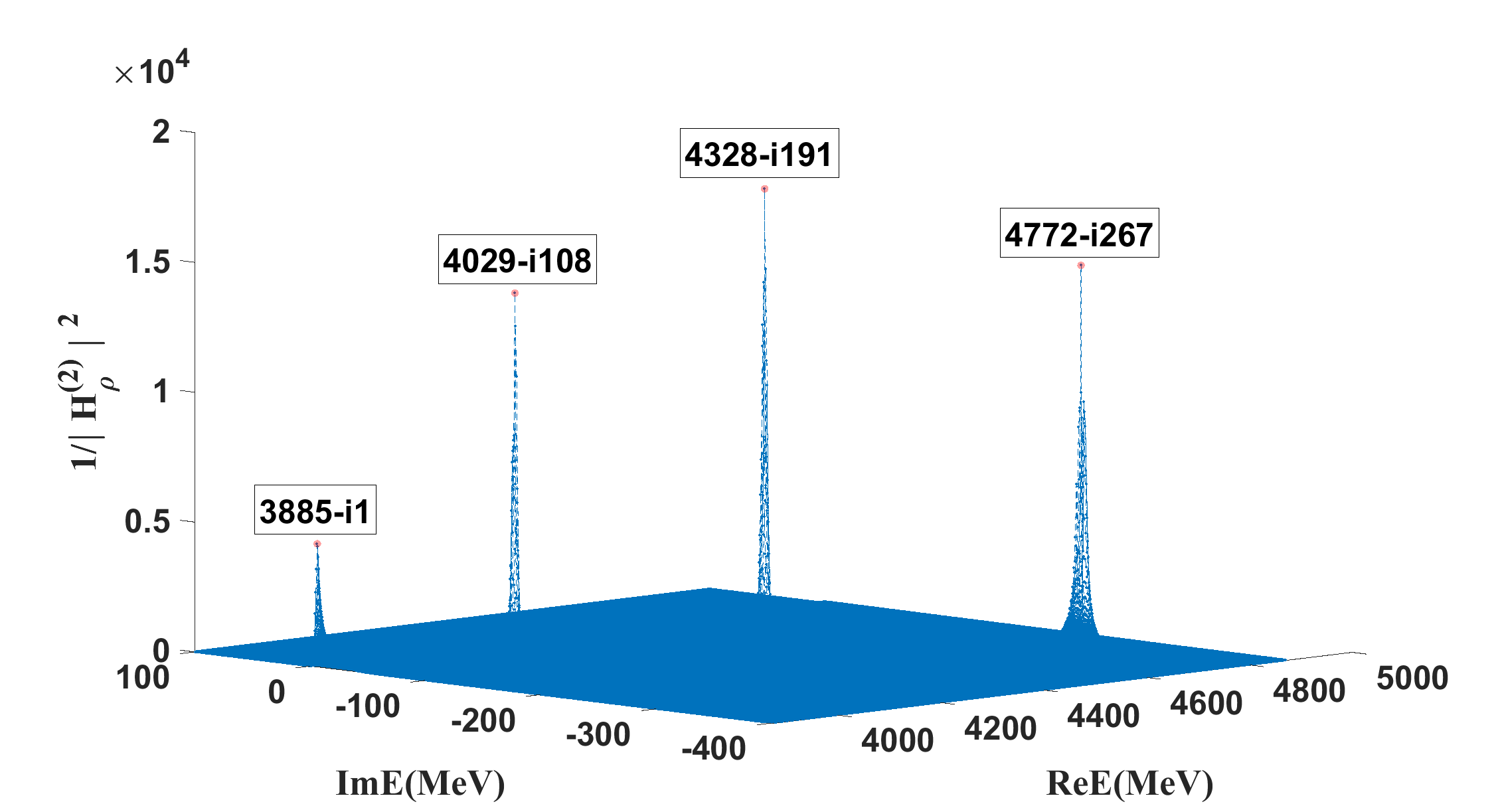}
\caption{$1/|H_\rho^{(2)}(\alpha)|^2$ .vs. the complex energy $E$ with $\alpha=2.405$ in the $D \bar{D}^*$ case. The pole of $1/|H_\rho^{(2)}(\alpha)|^2$ corresponds to a zero-point of the second kind of Hankel function $H_\rho^{(2)}(\alpha)$, which represents a $D^0 \bar{D}^{*0}$ resonance state, as labeled in the figure.}
\label{fig:ddstar}
\end{figure}

Altogether there are four resonance states generated by solving Eq.~(\ref{eq:outgoing}), as depicted in Fig.~\ref{fig:ddstar}, the first one lies at $3885-i1$MeV on the complex energy plane, which is higher than the $D \bar{D}^*$ threshold and might correspond to the $Zc(3900)$ particle. That is to say, if the  $ X(3872)$ particle is a bound state of $D$ and $\bar{D}^*$ mesons, the $Zc(3900)$ particle would be a resonance state of $D \bar{D}^*$.
In the calculation of this work, the isospin of the state can not be distinguished, therefore, all states generated dynamically are isospin degenerate.

In addition to the resonance state at $3885-i1$MeV, there are other three resonance states generated dynamically at $4029-i108$MeV, $4328-i191$MeV and $4772-i267$MeV as listed in Table.~\ref{Table:DDstar}, and it's  more possible that the last two resonance states correspond to the $\chi_{c1}(4274)$ and $\chi_{c1}(4685)$ particles, respectively.
Although the mass of the state at $4029-i108$MeV is close to the $\psi(4040)$ particle, this state owns a positive parity, while the parity of the $\psi(4040)$ particle is negative.
The $D \bar{D}^*$ system is investigated in the framework of the constituent quark model\cite{Kalashnikova:2005ui,Ortega:2009hj,Zhou:2017dwj,Deng:2023mza}, Lattice QCD\cite{Li:2024pfg} and the $D D^*$ interaction\cite{Giacosa:2019zxw,Wang:2023ovj},respectively, and a partner state of $X(3872)$ with $J^{PC}=1^{++}$ is predicted in these articles, which might correspond to the $X(3940)$ particle listed in the PDG data\cite{PDG}. In our calculation, a $D \bar{D}^*$ resonance state appears at $4029-i108$MeV on the complex energy plane, which might correspond to the partner state of the $X(3872)$ particle mentioned in these articles.
Anyway, we have given a hint to look for this possible resonance state by the experimental collaboration in future.
Consequently, it is concluded that all these four resonance states are radial excitation states of $D \bar{D}^*$ in the S-wave approximation.

\begin{table}[htbp]
 \renewcommand{\arraystretch}{1.2}
\centering
\vspace{0.5cm}
\begin{tabular}{c|c|c|c|c|c}
\hline\hline
 $D \bar{D}^*$ & energy &   name & $I^G(J^{PC})$ & mass & width \\
 \hline
1 & $3885-i1$ & $Zc(3900)$ & $1^+(1^{+-})$ &  $3887.1\pm2.6$  & $28.4\pm2.6$
  \\
2 & $4029-i108$  & $X(3940)$ & $?^?(?^{??})$    & $3942\pm9$ & $37^{+27}_{-17}$ \\                       
3 & $4328-i191$  &  $\chi_{c1}(4274)$ & $0^+(1^{++})$ &   $4286^{+8}_{-9}$ & $51\pm7$  \\                           
4 & $4772-i267$  & $\chi_{c1}(4685)$ &  $0^+(1^{++})$ &   $4684\pm7^{+13}_{-16}$ & $126\pm15^{+37}_{-41}$  \\
\hline\hline
\end{tabular}
\caption{
The complex energy of the possible $D \bar{D}^*$ ($\bar{D} D^*$) resonance state with the outgoing wave condition in Eq.~(\ref{eq:outgoing}) and the possible counterpart in the PDG data. All are in units of MeV.
}\label{Table:DDstar}
\end{table}

\begin{figure}
\includegraphics[width=0.8\textwidth]{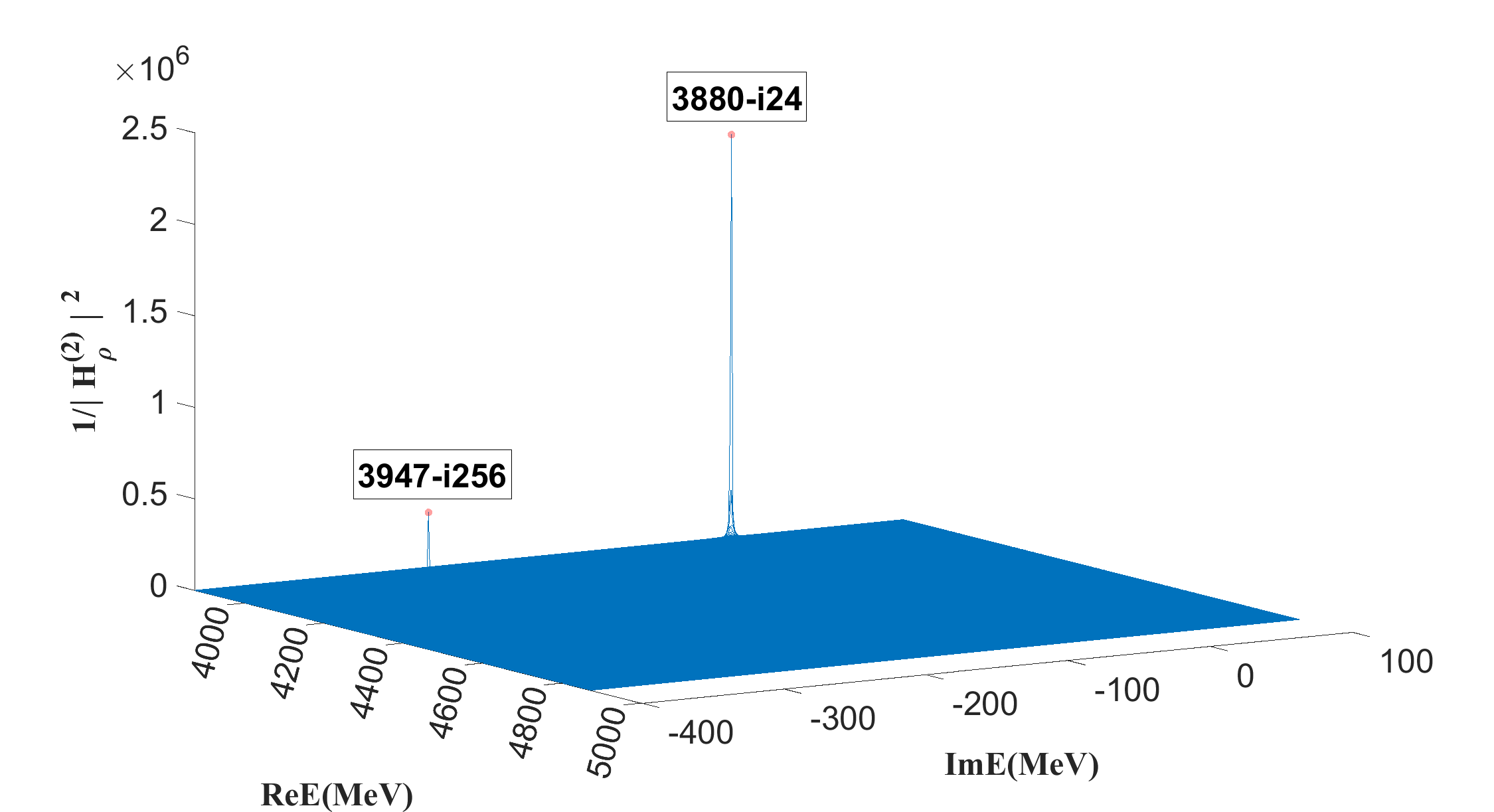}
\caption{$1/|H_\rho^{(2)}(\alpha)|^2$ .vs. the complex energy $E$ with $\alpha=4.8583$ in the $D \bar{D}^*$ case. The pole of $1/|H_\rho^{(2)}(\alpha)|^2$ corresponds to a zero-point of the second kind of Hankel function $H_\rho^{(2)}(\alpha)$, which represents a $D^{-} D^{*+}$ resonance state, as labeled in the figure.}
\label{fig:3dim}
\end{figure}

Since the $X(3872)$ particle almost lies at the $D^0 \bar{D}^{*0}$ threshold, the $X(3872)$ particle has been treated as a $D \bar{D}^*$ bound state and the binding energy is set to zero. Therefore, the zero point of the zeroth order Bessel function is evaluated in the calculation, which is relevant to the coupling constant in the $D^0 \bar{D}^{*0}$ Yukawa potential. However, the $D^{-} D^{*+}$ and $D^{+} D^{*-}$ channels also contribute to the production of $X(3872)$\cite{SunX3872}, which is about 8.11MeV lower than the $D^{-} D^{*+}$ or $D^{+} D^{*-}$ threshold\cite{PDG}, so the $X(3872)$ particle can also be regarded as a $D^{-} D^{*+}$ or $D^{+} D^{*-}$ bound state with a binding energy of 8.11MeV. Along this clue, the order of the Bessel function in Eq.~(\ref{eq:Besselzeropoint}) takes a value of 1.796, and the first zero point of the corresponding Bessel function lies at $\alpha=4.8583$. Therefore, the $D^{-} D^{*+}$ coupling constant in the Yukawa potential can be determined according to Eq.~(\ref{eq:couplingzeropoint}), i.e., $g=0.6520$, which is about twice of the $D^0 \bar{D}^{*0}$ coupling constant.
With the zero point of $\alpha=4.8583$, the eigenvalue of the Hamiltonian can be obtained according to Eq.~(\ref{eq:outgoing}). Consequently, two resonance states are generated as solutions of the Schrodinger equation. which lies at $3880-i24$MeV and $3947-i256$MeV on the complex energy plane, as shown in Fig.~\ref{fig:3dim}.
It is reasonable to assume the resonance state at $3880-i24$MeV correspond to the $Z_c(3900)$ particle, while the resonance state at
$3947-i256$MeV represent the $X(3940)$ particle in the PDG data\cite{PDG}. In addition, no other solutions are found at the higher energy region.
Although the binding energy becomes larger when the $X(3872)$ particle is treated as a $D^{-} D^{*+}$ bound state, it indicates that
the decay width of the resonance state does not decrease inevitable with the binding energy of the bound state increasing.

\section{Summary}
\label{sect:summary}

The $K \bar{ K}^*$ and $D \bar{ D}^*$ systems are studied by solving the Schrodinger equation under different boundary conditions of the wave function.
By fitting the binding energy of the $f_1(1285)$ particle, which is regarded as a $K \bar{K}^*$ or $\bar{K} { K}^*$ bound state in this work, the coupling constant in the $K \bar{ K}^*$ interaction is determined.
Sequently, a $K \bar{ K}^*$ resonance state is generated as a solution of the Schrodinger equation under the outgoing wave condition, which might correspond to the $f_1(1420)$ particle in the PDG data.
Similarly, this method is extended to study the $D \bar{ D}^*$ interaction. Supposing that the $X(3872)$ is a $D \bar{ D}^*$ bound state, several resonance states are obtained as solutions of the Schrodinger equation when the outgoing wave condition is taken into account.
Therefore, the relation between the bound state and the corresponding resonance state is established by solving the Schrodinger equation.

Especially, it is found that the one-pion exchanging potential plays an important role in order to produce the $K \bar{ K}^*$ and $D \bar{ D}^*$ resonance states, while the interaction via a vector meson exchanging is excluded in the calculation.
The eigenenergy of the Hamiltonian can be complex when the outgoing wave condition is taken into account, which implies that the inelastic scattering process is a non-Hermitian problem actually. Thus it would be an important method to study the properties of hadronic resonance states.

In summary, the $K \bar{ K}^*$ and $D \bar{ D}^*$  systems are studied by solving the Schrodinger equation with different boundary conditions. It is found that the one-pion-exchange potential plays an important role in the interaction of these two systems. By fitting the coupling constant with corresponding binding energy of the bound state, some resonance states are generated dynamically when the outgoing wave condition is taken into account, and most of them have a counterpart in the PDG data. Therefore, the calculation results manifest that there are intrinsic relations between the bound state and  resonance states of the system.


\end{document}